\documentclass[aps,preprint]{revtex4-1}%
\usepackage{amsfonts}
\usepackage{amsmath}
\usepackage{amssymb}
\usepackage{graphicx}
\usepackage{epstopdf}
\usepackage{epsfig}%
\providecommand{\U}[1]{\protect\rule{.1in}{.1in}}

\newcommand{\be}{\begin{equation}}
\newcommand{\ee}{\end{equation}}
\newcommand{\bea}{\begin{eqnarray}}
\newcommand{\eea}{\end{eqnarray}}
\newcommand{\ba}{\begin{array}}
\newcommand{\ea}{\end{array}}

\begin{document}
\title{Monogamy of Information Causality }
\author{Li-Yi Hsu}
\affiliation{${}$Department of Physics, Chung Yuan Christian University, Chung-li 32023, Taiwan}

\begin{abstract}
We consider information causality in the multi-receiver random access codes.
Therein, no receiver can gain any information only from classical
communication. We claim the following statement. Information causality still
holds with the help of the multi-partite physical non-local resource. That is,
the summation of all revivers' information gain cannot be greater than the
amount of classical communication. The distributive multi-party physical
nonlocal resource can be exploited only for information splitting. It is
proved that such trade-off leads to the monogamy of entanglement. Finally the
connection between information causality and spin-glass Bethe lattice is discussed.

\end{abstract}
\startpage{1}
\endpage{2}
\maketitle

For almost a hundred years, quantum theory has been a successful theory in
describing microscopic physics. Quantum theory is essentially grounded on the
mathematical axioms. Recently attempts have been made to reformulate quantum
physics using physical principles. In contrast to classical physics, quantum
theory can be featured by no-cloning, secret privacy, no-broadcasting,
however, which are common among generic nonlocal theories \cite{03,04}.
However, so far the only nonlocal theory that is physically realized is
quantum theory \cite{1,2}. There should be a profound physical principle. Such
principle can be exploited as the basis of quantum theory. It can also single
out quantum theory as a physical theory.

In the hunting for the hidden principle, two potential candidates, macroscopic
locality \cite{ML}\ and information causality \cite{IC}, have been proposed
recently. To simply put, macroscopic locality states that the Bell-type
experiment with a sufficient number of detectors can falsify quantum
mechanics. In other words, the statistics of coarse-grained outcome
correlation should admit a local hidden variable model. On the other hand,
information causality states that one can gain no more than $m$-bit
information of the other's database, once $m$-bit classical communication is
allowed. Notably, even in the bipartite scenario, macroscopic locality and
information causality are inequivalent with multi-level qudits \cite{nc}. As
for the fulfillment of macroscopic locality, the global covariance matrix
should be tested semidefinite positive.

In information processing, no-signaling correlation as physical resource can
be quite useful. For example, communication complexity can be trivial if a
physical theory were to allow maximal nonlocal correlation \cite{5,6}.\ From
the information theoretic viewpoint, a relevant question can be stated as
follows. Why can quantum theory not be more nonlocal? Historically, Popescu
and Rohrlich demonstrated that maximal nonlocality can be achieved without
violating the no-signaling principle, due to the request of relativity
\cite{pr}. In the bipartite communication protocol, according to the
no-signaling principle, distant Alice and Bob each cannot gain the other's
information only with accessible nonlocal resource. With classical
communication, nonlocality can help gain the other's information. Intuitively,
one can gain more information if the resource were more nonlocal. As a
physical principle, information gain is upper-bounded by information causality
\cite{IC}. In this way, a nonlocal theory can be calibrated into a physical
one, i.e. quantum theory, using information causality. Any theory more
nonlocal than the quantum theory must violate information causality and hence
be unphysical.

In this Letter, we investigate information causality in the framework of
distributive $(n,k)$ random access codes ($(n,k)$-RAC) \cite{rac}. Wherein
Alice can directly access a local $k$-bit database, which comprises the
components of the vector $\overrightarrow{a}:=$ ($a_{0}$, $a_{1}$, $\cdots$
$a_{k-1}$), with $a_{i}$ being the random variable and $a_{i}$ $\in\{0,1\}$
$\forall i$. There are $n$ distant parties, Bob$_{1}$, Bob$_{2}$,$\cdots$,
Bob$_{n}$. $\forall j$, Bob$_{j}$ is given the random variable $b_{j}%
\in\{0,1,\cdots,k-1\}$, and his task is to optimally guess $a_{b_{j}}$. In the
following Alice is allowed to public announce a bit, $c$, via classical
communication. Denote Bob$_{j}$'s local information gain as $I_{j}=%
{\displaystyle\sum_{l=0}^{k-1}}
I(a_{l}:\beta_{j}|b_{j}=l)$, where $I(a_{l}:\beta_{j}|b_{j}=l)$ is the Shannon
mutual information between $a_{l}$ and Bob$_{j}$'s guessing answer $\beta_{j}$
under the condition that Bob$_{j}$ has received $b_{j}=l$. Ideally,
$I(a_{l}:\beta_{j}|b_{j}=l)=1$ if $a_{l}=\beta_{j}$ always holds. That is,
Bob$_{j}$ always guesses $a_{l}$ correctly. The information causality quantity
is defined as
\begin{equation}
I=%
{\displaystyle\sum_{j=1}^{n}}
I_{j} \label{total}%
\end{equation}

The $n=1$\ case has been fully studied in \cite{IC,hsu}. Therein, information
causality can be stated as follows. Under the limit of the one-bit
broadcasting, ,the nonlocal resource can be physically realized, if
\begin{equation}
I\leq1, \label{IIC}%
\end{equation}
Notably, the above inequality can be realized either classically or quantumly.
However, it is not the case once $n\geq2$. For example, Alice can send the bit
$c=a_{0}$ and Bob$_{j}$'s answer $\beta_{j}$ is always equal to $c$. Hence
$I(a_{l}:\beta_{j}=c|b_{j}=l)$ $=\delta_{l,0}$ and $I=n$. In this case, $I$ is
an increasing function of $n$ and goes infinite as $n\rightarrow\infty$. In
the following it is required that $I(a_{l}:c)$ $=0$ $\forall l$. The reasons
are twofold. Firstly, any receiver gains no information from the communication
bit. The non-zero information gain must intrinsically come from the accessible
nonlocal resource. Secondly, in cryptography, if the accessibility of the
database is restrictive to the legitimate users, no information leakage should
be allowed via classical communication. Eventually, if there is no nonlocal
resource distributed between distant parties, $I=0$.

To perform the $(n,k)$-RAC,\ Alice and Bob$_{j}$ $\forall j$ each input some
bits into the no-signaling boxes, which then output a bit. The output
correlation can be either local or nonlocal without violating the no-signaling
principle. In this Letter, we claim the monogamy of information causality as
follows. \emph{With broadcasting one classical bit carrying no information,
the nonlocal boxes can be physically realized if the condition }(\ref{IIC})
\emph{still holds in the multi-receiver scenario}. In other words, even
exploiting multi-partite entangled states cannot gain more information than
one bit. Rather, nonlocal resource can only split information into different receivers.

The monogamy or shareability of entanglement or nonlocal correlation has been
studied about a decade. Coffman et al considered the trade-off relation of
three-qubit system using the measure of bipartite entanglement known as tangle
\cite{m2}. Later such trade-off relation of $N$-qubit system was studied
by\ Osborne and Verstraete \cite{m21}. The limited shareability of generic
correlations was also studied \cite{m1,m5}. Koashi and Winter considered the
trade-off between quantum and classical correlations \cite{m3}. Notably, these
monogamy relations are linear. Toner and Verstraete originally considered the
monogamy relation of three-qubit system from the aspect of bipartite
nonlocality \cite{bf}. The summation of the quadratic
Clauser-Horne-Shimony-Holt (CHSH) values \cite{chsh} is limited. Recently
Kurzy\'{n}ski et al investigated the monogamy relation of $N$-qubit system
using multipartite nonlocality \cite{m6}. Therein the summations of the
quadratic Seevenick-Bell \cite{sb}\ or Mermin-Bell \cite{mermin} values are
also limited. In this Letter, it will be shown that information monogamy is
deeply connected with that of entanglement \cite{bf,m6}. Specifically the
monogamy relations of three- and four-qubit entanglement can be directly
derived using the monogamy of information causality in the $(3,2)$-RAC.
Consequently monogamy of entanglement can be operationally meaningful from the
information theoretic viewpoint. In \cite{m6}, quadratic monogamy of
entanglement is studied with two measurement settings on each qubit. Such
trade-off relation is essentially related to local realistic description of
the correlation functions, which lead to $%
{\displaystyle\sum_{k_{1}\ldots k_{N}=x,y}}
T_{k_{1}\ldots k_{N}}^{2}$ $\leq1$, \ where $T_{k_{1}\ldots k_{N}}=\frac
{1}{2^{N}}Tr(\rho\sigma_{k_{1}}\otimes\ldots\otimes\sigma_{k_{N}})$, $\rho$ is
the $N$-qubit density matrix, and, under local unitary, $\sigma_{k_{i}}$ can
be set as the Pauli operator either $\sigma_{x}$ or $\sigma_{y}$ for $i$-th
qubit \cite{tt}. In this Letter the multi-setting monogamy of entanglement can
be also derived using Ineq. (\ref{IIC}).

On the other hand, monogamy of information causality can guarantee the
unconditional security in quantum cryptography. Actually, the maximal $I$
depends on the amount of classical communication, rather than the number of
nonlocal boxes. Suppose the one-bit sender Alice wants to share the secret
bits with Bob$_{1}$ via quantum channels. Whereas Bob$_{2}$, Bob$_{3}$,
$\cdots$Bob$_{n}$ are regarded as collaborative eavesdroppers. Once
$I_{1}>1/2$ $>%
{\displaystyle\sum_{j=2}^{n}}
I_{j}$, using any attack equipped with unlimited physical multipartite
nonlocal resource, these eavesdroppers can gain no more information than the
receiver. Quantumly, $I_{1}>1/2$ can be done if Alice and Bob$_{1}$ share more
entanglement than that shared between Alice and all other Bob$_{j}$s
\cite{science}. With afterward error correction and privacy amplification
\cite{qc}, Bob$_{1}$ can access these bits in secure. Here we take BB84
protocol as a simple example with $n=2$. Therein, Bob$_{1}$ and Bob$_{2}$ are
the receiver and the eavesdropper, respectively. Once the sender Alice
announces 1-bit information of preparation basis, which is irrelevant to the
secret bit. It is straightforward that $I_{2}=-Q\log_{2}Q-(1-Q)\log
_{2}(1-Q)=1-I_{1}$, where $Q$ is the quantum bit error rate. As a result,
$I=1$.

As a final remark, the bipartite nonlocal correlation exploited in the
$(1,k)$-RAC has been fully studied \cite{hsu}. Therein Ineq. (\ref{IIC}) can
be reduced as a convex optimization problem \cite{wehner}. Hence information
causality can be numerically validated by semidefinite programming \cite{hsu,
sdp}. However, such reduction is unknown in general $(n,k)$-RAC, since
multipartite correlation is involved. Eventually, as a physical hypothesis,
monogamy of information causality must be verified or falsified only by
physical experiments. Hereafter, the addition is the addition modulo two.

$(n,2)$\textit{-RAC} --- \ Before further proceeding, $(1,2)$\textit{-RAC} is
reviewed as follows. Firstly Alice inputs the bit $x=a_{0}+a_{1}$\ into the
box which outputs $A$; Bob$_{1}$ inputs the bit $y_{1}=b_{1}$ into the
accessible box which outputs $B_{1}$. Alice sends the bit $c=a_{0}+A$ and
Bob$_{1}$'s guess answer is $\beta_{1}=c+B_{1}$. As for its physical
realization, two entangled qubits are exploited as a\ nonlocal box (NL-box).
The inputs and outputs correspond to the measurement\ settings and the
outcomes, respectively \cite{IC,tt}. In the $(n,2)$-RAC case, similar to
Bob$_{1}$'s processing, Bob$_{j}$ inputs the bit $y_{j}=b_{j}$ into the
accessible box which outputs $B_{j}$. Then Bob$_{j}$'s guessing answer is
$\beta_{j}=c+B_{j}$. If the unphysical Popescu-Rohrlich boxes are exploited
such that, in the bipartite scenario, $A+B_{j}=xy_{j}$. Bob$_{i}$ can access
both $a_{0}$ and $a_{1}$ and hence $I_{i}(a_{l}:\beta_{j}|b_{j}=l)=1$ $\forall
l\in\{0,1\}$. Given $b_{j}$, let the $B_{j}=xy_{j}+A$ with probability
$\frac{1}{2}(1+\xi_{j,l})$. Here Bob$_{i}$'s box is regarded a gate that errs
with the probability $\frac{1}{2}(1-\xi_{j,l})$. That is,
\begin{equation}
\xi_{j,l}=P(A+B_{j}=xy_{j})-P(A+B_{j}=xy_{j}+1). \label{cau}%
\end{equation}
\ Before proceeding further Evans-Schulman lemma should be introduced as
follows \cite{es,es2}.

\textit{Lemma :} Let $X$ and $Y$ be random variables. Let the binary symmetric
channel $C$ be
\begin{equation}
\left(
\begin{array}
[c]{cc}%
\frac{1+\xi}{2} & \frac{1-\xi}{2}\\
\frac{1-\xi}{2} & \frac{1+\xi}{2}%
\end{array}
\right)  \text{.} \label{channel}%
\end{equation}
Let $Y$ and $Z$ be the random variable input and output of the symmetric
channel $C$, respectively. Let the random variables $Q$ such that $Z$ is
independent of ($Q$, $X$). Then
\[
\frac{I(X;Z|Q)}{I(X;Y|Q)}\leq\xi^{2}.
\]

\textit{Proof:} See Ref. \cite{es,es2} for the detailed rigorous proof.
Therein, $Q$\ is not necessarily binary.

In the following Bob$_{j}$'s noisy box is decomposed as the perfect box
attached with the noisy channel $C$. Specifically, the output of perfect box,
$xy_{j}+A$, is the input of the channel, which outputs $B_{j}$. According to
(\ref{channel}), $B_{j}$ is equal to $xy_{j}+A$ with probability $\frac{1+\xi
}{2}$ with $\xi=\xi_{j,l}$. Such the disturbing noise in the physical boxes is
intrinsic and hence unavoidable in the physical implementation. Regarding
Bob$_{j}$'s information gain, $Q$\ is set as the given condition $b_{j}=l$,
and $X$ is set as $a_{l}$. $Y$ is set as the addition of the communication bit
$c$ and the output of the perfect box. In other words, $Y=c+(xy_{j}+A)=a_{l}$
and $I(a_{l};Y|b_{j}=l)=1$. Finally $Z=c+B_{j}$\ and with the flipping
probability equal to $\frac{1}{2}(1-\xi_{j,l})$. Consequently we have
\[
I(a_{k};\beta_{i}|b_{i}=l)\leq\xi_{j,l}^{2},
\]
and hence $I_{j}\leq%
{\displaystyle\sum_{l=0}^{k-1}}
\xi_{j,l}^{2}$. For the quantum correspondence, the inputs $x$ and
$y_{j}(=b_{j})$ correspond to physical observables $\widehat{A}_{x}$ and
$\widehat{B}_{j,y_{j}}$, respectively, and the outputs to the measurement
outcomes. Once the output 0 and 1 are mapped into 1 and -1, according to
(\ref{cau}), we have $\xi_{j,l}=\frac{1}{2}%
{\displaystyle\sum_{x=0,1}}
(-1)^{xl}\left\langle \widehat{A}_{x}\widehat{B}_{j,l}\right\rangle $. Using
Cauchy--Schwarz inequality and denoting $\mathcal{CHSH}_{AB_{j}}=2(\xi
_{j,0}+\xi_{j,1})$ \cite{chsh}, we have
\begin{equation}%
{\displaystyle\sum_{j=1}^{n}}
\mathcal{CHSH}_{AB_{j}}^{2}\leq8.\label{mo1}%
\end{equation}
Notably the $n=2$ case is exactly equal to the monogamy of three-qubit
entanglement \cite{bf}.

$(n,2)$-RAC can be alternatively processed as follows. Alice exploits two
boxes, where $x_{1}=a_{0}$ and $x_{2}=a_{1}$ are the inputs and $A_{1}$ and
$A_{2}$ are the outputs respectively. Then Alice sends the bit $c=A_{1}+$
$A_{2}+a_{0}+\overline{a_{1}}+a_{0}\overline{a_{1}}$. Similarly, Bob$_{j}$
input $y_{j}=b_{j}$ to the box that outputs $B_{j}$. Finally $\beta_{j}%
=B_{j}+c$ if $y_{j}=0$ and $\beta_{j}=B_{j}+c+1$ if $y_{j}=1$. (Another very
similar $(1,k)$-RAC has been recently proposed \cite{china}.) Notably, if
$A_{1}+$ $A_{2}+B_{j}=x_{1}x_{2}y_{j}+\overline{x_{1}}$ $\overline{x_{2}}$
$\overline{y_{j}}$, once the output 0 and 1 are mapped into 1 and -1,
respectively, the Seevinck-Bell value $\mathcal{SB}=$ $%
{\displaystyle\sum_{x_{1},x_{2},y_{j}=0}^{1}}
(-1)^{x_{1}x_{2}y_{j}+\overline{x_{1}}\overline{x_{2}}\overline{y_{j}}%
}\left\langle x_{1}x_{2}y_{j}\right\rangle $ can be saturated up to 8, whereas
its Tsirelson bound is $4\sqrt{2}$. Quantumly, the corresponding physical
observables of inputs $x_{1}$and $x_{2}$ are $\widehat{A}_{x_{1}}$ and
$\widehat{A}_{x_{2}}^{\prime}$, respectively, and as a result, $\xi
_{j,0}=\frac{1}{4}%
{\displaystyle\sum_{x_{1},x_{2}=0}^{1}}
(-1)^{\overline{x_{1}}\text{ }\overline{x_{2}}}\left\langle \widehat{A}%
_{x_{1}}\widehat{A}_{x_{2}}^{\prime}\widehat{B}_{j,0}\right\rangle $,
$\xi_{j,1}=\frac{1}{4}%
{\displaystyle\sum_{x_{1},x_{2}=0}^{1}}
(-1)^{x_{1}x_{2}}\left\langle \widehat{A}_{x_{1}}\widehat{A}_{x_{2}}^{\prime
}\widehat{B}_{j,1}\right\rangle $ and hence $\mathcal{SB}_{A_{1}A_{2}B_{j}%
}=4(\xi_{j,0}+\xi_{j,1})$. Then we have
\begin{equation}%
{\displaystyle\sum_{j=1}^{n}}
\mathcal{SB}_{A_{1}A_{2}B_{j}}^{2}\leq32\label{4q}%
\end{equation}
For $n=2$\ case with exchanging Alice' boxes with Bob$_{1}$ and Bob$_{2}$'s
boxes, we have
\begin{equation}
\mathcal{SB}_{A_{1}A_{2}B_{1}}^{2}+\mathcal{SB}_{A_{1}A_{2}B_{2}}%
^{2}+\mathcal{SB}_{B_{1}B_{2}A_{1}}^{2}+\mathcal{SB}_{B_{1}B_{2}A_{2}}^{2}%
\leq64.\label{SB}%
\end{equation}
An alternative form is Mermin three-qubit monogamy $\mathcal{M}_{A_{1}%
A_{2}B_{1}}^{2}+\mathcal{M}_{A_{1}A_{2}B_{2}}^{2}+\mathcal{M}_{B_{1}B_{2}%
A_{1}}^{2}+\mathcal{M}_{B_{1}B_{2}A_{2}}^{2}\leq16$ \cite{m6}, where
three-qubit Mermin-Bell value $\mathcal{M}_{A_{1}A_{2}B_{j}}=$ $\left\langle
\widehat{A}_{1}\widehat{A}_{0}^{\prime}\widehat{B}_{j,0}\right\rangle
+\left\langle \widehat{A}_{0}\widehat{A}_{1}^{\prime}\widehat{B}%
_{j,0}\right\rangle +\left\langle \widehat{A}_{0}\widehat{A}_{0}^{\prime
}\widehat{B}_{j,1}\right\rangle -\left\langle \widehat{A}_{1}\widehat{A}%
_{1}^{\prime}\widehat{B}_{j,1}\right\rangle $ \cite{mermin}. Using
Cauchy-Schwarz inequality and permuting the subscripts 0$\leftrightarrow$1,
Mermin monogamy can be lead to Ineq. (\ref{SB}).

Before further proceedings of $(n,k)$\textit{-RAC}, here we argue the validity
of the monogamy of information causality. In the original proposal of
information causality, the bipartite protocol of $(1,k)$\textit{-RAC }is
studied. Therein, $b$ is given to Bob to guess $a_{b}$. Since the
multi-partite nonlocal resource is accessible, now Bob is required to guess
$n$\ elements $a_{b_{1}}$,\ldots$a_{b_{n}}$ , of $\overrightarrow{a}$. As for
Bob, he divides the boxes in $n$ groups. Those boxes in $i$-th group are
exploited for guessing $a_{b_{i}}$. Notably, if $\ n>k$, there must be some
$p$ and $q$, such that $b_{p}=b_{q}$. Bob can exploit the boxes\ in the $p$-th
and $q$-th groups jointly to guess out $a_{b_{p}}$. According to definition of
$I$ and information causality, Bob's information gain is no more than one bit.
As for the multi-partite protocol of $(n,k)$\textit{-RAC}, boxes of different
groups are spatially separated. No other extra classical-bit broadcast from
Bob$_{i}$ $\forall$ $i$ is allowed. In this condition, the value $I$ cannot be
increased. Hence the monogamy of information causality should be guaranteed.

$(n,k)$\textit{-RAC --- }The protocol is similar that of $(1,k)$\textit{-RAC
}as follows. Alice inputs the ($k-1$)-bits $\alpha_{1},\ldots,\alpha_{k-1}$,
where $\alpha_{i}=a_{0}+a_{i}$, into the box. Bob$_{j}$ input the random
variable $b_{j}$ into the box that outputs $B_{j}$. After Alice announces the
bit $c=a_{0}+A$, Bob$_{j}$'s guess answer $\beta_{j}=c+B_{j}=a_{0}+A+B_{j}$.

Let the $k$-bit vector $\overrightarrow{\alpha}=(\alpha_{0}\ldots\alpha
_{k-1})$ with $\alpha_{0}$ always being $0$. Assuming that $b_{j}=j^{\prime}$,
let another $k$-bit vector $\overrightarrow{b_{j}}=(b_{j,0}\ldots b_{j,k-1})$,
where $b_{j,0}=\delta_{j,j^{\prime}}$. Notably the Hamming weight of
$\overrightarrow{b_{j}}$ is always 1. As a result, $a_{b_{j}}=\beta_{j}$ if
$A+B_{j}=$ $\overrightarrow{\alpha}\cdot$ $\overrightarrow{b_{j}}%
=a_{0}+a_{j^{\prime}}$. Quantumly, the corresponding observables of
$\overrightarrow{\alpha}$\ and $\overrightarrow{b_{j}}$ are $\widehat
{M}_{\overrightarrow{\alpha}}$ and $\widehat{N}_{\overrightarrow{b_{j}}}$,
respectively. With straight calculation,
\begin{equation}
\xi_{j,\overrightarrow{b_{j}}}=\frac{1}{2^{k-1}}%
{\displaystyle\sum_{\{\overrightarrow{\alpha}\}}}
(-1)^{\overrightarrow{\alpha}\cdot\overrightarrow{b_{j}}}\left\langle
\widehat{M}_{\overrightarrow{\alpha}}\widehat{N}_{\overrightarrow{b_{j}}%
}\right\rangle
\end{equation}
and hence
\begin{equation}
I_{j}=%
{\displaystyle\sum_{\{\overrightarrow{b_{j}}\}}}
\xi_{j,\overrightarrow{b_{j}}}^{2}. \label{gain}%
\end{equation}

Here we consider the $n=1$ case. Denote $\mathcal{IC}_{AB_{j}}=%
{\displaystyle\sum_{\{\overrightarrow{\alpha},\overrightarrow{b_{j}}\}}}
(-1)^{\overrightarrow{\alpha}\cdot\overrightarrow{b_{j}}}\left\langle
\widehat{M}_{\overrightarrow{\alpha}}\widehat{N}_{\overrightarrow{b_{j}}%
}\right\rangle $ as the Bell value. Again, using the Cauchy-Schwarz inequality
and straight calculation, a nonlocal physical theory must obey the following inequality%

\begin{equation}
\sum_{m=0}^{k-1}\binom{k-1}{m}\left\vert k-2m\right\vert \leq|\mathcal{IC}%
_{AB_{1}}|\leq2^{k-1}\sqrt{k}. \label{main}%
\end{equation}
Both local realism and superquantum correlations each violate the above
inequality in the different ways. As for arbitrary $n$, Ineqs. (\ref{IIC})and
(\ref{gain}) lead to the following trade-off relation
\[%
{\displaystyle\sum_{j=1}^{n}}
\mathcal{IC}_{AB_{j}}^{2}\leq4^{k-1}k.
\]

Notably, an alternative processing of ($1$, $k$)-RAC has been proposed by
Paw\l owski et al \cite{IC}. For simplicity, set $k=2^{p}.$ Notably, the
nonlocal boxes accessible for each receiver\ are locally exploited as noisy
gates comprising the circuit $G$, which corresponds to the $2$-ary $p$-depth
complete tree. Here we can regard $G$ as Bethe lattice with open boundary
\cite{bethe}. The spin value $s_{i}$\ for each vertex $i$\ is assigned as
1($-1$)\ if the corresponding box outputs 1(0). As a result the spin
configuration is the ground state of the spin glass Ising model with
Hamiltonian
\begin{equation}
H=-%
{\displaystyle\sum\limits_{(i,\text{ }j)\in E}}
J_{ij}s_{i}s_{j}, \label{is}%
\end{equation}
where $(i,$ $j)\in E$ if vertex $i$ and $j$ are connected. At the zero
temperature ($T=0$), $s_{i}=s_{i}^{0}$ for the vertex $i$, which is the
noiseless output of the corresponding box. Artificially the coupling strength
$J_{ij}$ is set as $s_{i}^{0}s_{j}^{0}J$, where $J>0$. Hence $J=\pm J$ with
probability $\frac{1}{2}$, and the spin configuration at $T=0$ composes the
ground state. The random noise of the boxes corresponds to the thermal
fluctuation. Consequently $\xi$ in (\ref{channel}) is equal to $|<s_{i}>|$,
which is decreased with increasing $T$, and $n$. The lattice $G$ is
paramagnetic and all boxes are local at $T\rightarrow\infty$. Here we
conjecture as follows. The lattice $G$ comprised by physical nonlocal boxes
are always \emph{paramagnetic} and never be in spin glass or ferromagnetic
phase \cite{b1}. In other words, $I=1$ may imply the occurrence of phase
transition in the Bethe lattice $G$.

The authors acknowledge national support from the National Science Council of
the Republic of China under Contract No. NSC.99-2112-M-033-007-MY3. This work
is partially supported by the Physics Division of the National Center for
Theoretical Sciences.

\end{document}